\documentclass[mathematics,article,accept,moreauthors,dvipdfm,12pt,a4paper]{mdpi}

\lastpage{x}
\doinum{10.3390/------}
\pubvolume{xx}
\pubyear{2015}

\history{Received: xx / Accepted: xx / Published: xx}

\usepackage{amssymb}
\usepackage{amsmath}
\usepackage{amscd}
\usepackage{latexsym}
\usepackage{graphicx}

\newcommand{\be}{\begin{equation}}
\newcommand{\ee}{\end{equation}}

\newcommand{\br}{{\bf r}}

\newcommand{\bt}{\beta}
\newcommand{\vp}{\varphi}
\newcommand{\ep}{\varepsilon}
\newcommand{\al}{\alpha}
\newcommand{\ra}{\rightarrow}

\newcommand{\om}{\omega}

\newcommand{\Gm}{\Gamma}

\Title{Effective Summation and Interpolation of Series by Self-Similar
Root Approximants}

\Author{Simon Gluzman and Vyacheslav I. Yukalov*}

\address{%
Bogolubov Laboratory of Theoretical Physics,
Joint Institute for Nuclear Research, \\
Dubna 141980, Russia}

\corres{yukalov@theor.jinr.ru, +7(496) 21 63 947}

\abstract{
We describe a simple analytical method for effective summation of series,
including divergent series. The method is based on self-similar
approximation theory resulting in self-similar root approximants.
The method is shown to be general and applicable to different problems,
as is illustrated by a number of examples. The accuracy of the method is
not worse, and in many cases better, than that of Pad\'{e} approximants,
when the latter can be defined.
}


\begin{document}

\section{Introduction}

In numerous cases of applied mathematics and mathematical physics the
solutions to problems can only be represented as series derived by means
of some kind of perturbation theory or iterative procedure. A great
majority of such series is even divergent, having meaning only as
asymptotic series for an infinitesimally small expansion variable. While
the considered problems often require to consider finite values of this
variable, sometimes even very large values. The standard way of treating
such asymptotic series, for the purpose of their extrapolation to the 
finite values of the variable, is by invoking the Pad\'{e} approximants 
\cite{Baker1}. The latter, however, exhibit several deficiencies limiting 
their applicability, as is discussed in Refs. \cite{Baker1,Gluzman2}, for 
instance, such a notorious deficiency as the appearance of spurious poles. 
Another weak point is the ambiguity of choosing one of the Pad\'{e} 
approximants $P_{M/N}$ from the table of many admissible, for each series 
of order $k$, variants satisfying the condition $M + N + 1 = k$. Also, in 
the limit of a large variable $x$ the approximant $P_{M/N}(x)$ behaves as 
$x^{M - N}$. Hence, only integer powers of $x$ are allowed. It is possible 
to improve the results by employing the modified Pad\'{e} approximants 
\cite{Baker_3}, corresponding to the power $P_{M/N}^\gamma$, with choosing 
the appropriate value of $\gamma$ satisfying the large-variable limit.

In the present paper, we show that it is possible to formulate a general
method for effectively extrapolating and interpolating asymptotic series. 
The method enjoys the following advantages: (i) It is unambiguously defined 
for each given series of order $k$; (ii) It allows for the treatment of 
large-variable behavior of any type, whether with integer, rational, or 
irrational powers; (iii) Being more general, it is not less accurate than 
the method of the Pad\'{e} approximants, when the latter exist, in many 
cases, being more accurate.

In the great majority of realistic situations, only a few terms of 
asymptotic expansions are available. Therefore, in the examples below, we 
do not consider very large series, showing that even several of terms allow 
us to derive quite accurate approximations.

\section{Self-similar root approximants}

Suppose we are interested in finding a real function $f(x)$ of a real 
variable $x$. However, this function is defined by a complicated equation 
that cannot be solved exactly. But, applying a kind of perturbation theory, 
we can derive the small-variable behavior of this function
\be
\label{1}
 f(x) \simeq f_k(x) \qquad ( x \ra 0 ) \;  ,
\ee
represented by asymptotic series, with the $k$-th order expansion
\be
\label{2}
 f_k(x) = f_0(x) \left ( 1 + \sum_{n=1}^k a_n x^n \right ) \;  ,
\ee
where
\be
\label{3}
 f_0(x) = A x^\al \;  .
\ee

Sometimes, the large-variable behavior of the function
\be
\label{4}
f(x) \simeq f^{(p)}(x) \qquad ( x \ra \infty )
\ee
is also known and can be represented by an expansion over $1/x$ as
\be
\label{5}
f^{(p)}(x) = f_\infty(x) \left ( 1 + \sum_{n=1}^p \; \frac{b_n}{x^n}
\right ) \;  ,
\ee
with
\be
\label{6}
 f_\infty(x) = B x^\bt \;  .
\ee

For what follows, it is convenient to deal with the ratio $f(x)/f_0(x)$,
which at small variable $x \ra 0$ behaves as
\be
\label{7}
 \frac{f(x)}{f_0(x)} \simeq \frac{f_k(x)}{f_0(x)} = 1 +
\sum_{n=1}^k a_n x^n \;  ,
\ee
and at large values of the variable $x \ra \infty$ it tends to
\be
\label{8}
  \frac{f(x)}{f_0(x)} \simeq \frac{f_\infty(x)}{f_0(x)} =
\frac{B}{A} \; x^{\bt-\al} \; .
\ee

The extrapolation of the small-variable expansions to the large-variable
region can be done by means of self-similar approximation theory
\cite{Yukalov_4,Yukalov_5,Yukalov_6,Yukalov_7,Yukalov_8}. In this approach,
the transfer from a $k$-th order approximation, say, a small-variable
expansion, to the higher orders of approximations are treated as the
motion with respect to the approximation order $k$ playing the role of
discrete time. Constructing a dynamical system, whose trajectory is
bijective to the sequence of approximations, makes it feasible to find a
fixed point representing the sought function. The convergence to the fixed
point is governed by control functions. The self-similar approximation 
theory combines the methods of optimal control theory, dynamical theory, 
and renormalization-group approach. We shall not go into the details and
mathematical justification of the self-similar approximation theory that 
has been thoroughly expounded in Refs.
\cite{Yukalov_4,Yukalov_5,Yukalov_6,Yukalov_7,Yukalov_8}, but we shall
use some of its consequences.

Employing this theory for the purpose of interpolation between the
small-variable and large-variable regions, it is possible to come
\cite{Yukalov_9,Gluzman_10,Yukalov_11} to the self-similar root approximant
\be
\label{9}
 \frac{f^*_k(x)}{f_0(x)} = \left ( \left ( \ldots ( 1 + A_1 x )^{n_1}
+ A_2 x^2 \right )^{n_2} + \ldots + A_kx^k \right )^{n_k} \; .
\ee
A theorem has been proved \cite{Yukalov_12} stating that all parameters
$A_i$ and powers $n_i$ of approximant (9) are uniquely defined through the
large-variable form (5).

However, the root approximant (9) cannot be uniquely defined through the
small-variable expansion (2). This hinders the applicability of the 
approximant (9), since in the majority of cases, the small-variable expansion 
is better known, providing a number of terms, while the knowledge of the 
large-variable behavior is limited by just a single term (6), often even 
without precise data for the amplitude $B$. In order to extend the 
applicability of approximant (9) to be uniquely defined through the 
small-variable expansion, it is necessary to impose some constraints on 
the powers $n_j$. Such a straightforward constraint is the requirement that 
all parameters $A_j$ of approximant (9) be involved in the definition of the 
large-variable limit, which implies the relation
\be
\label{10}
 n_j = \frac{j+1}{j} \qquad ( j = 1,2,\ldots, k-1) \;  ,
\ee
with $n_k = \beta - \alpha$. By expanding Eq. (9) in powers of $x$, it
is easy to prove that all parameters $A_j$ are uniquely defined through
the coefficients $a_j$ of small-variable expansion (2). In addition, we
can require the validity of the limiting form (6), which improves accuracy.

The self-similar root approximant (9), with conditions (6) and (10),
whose parameters $A_j$ are uniquely defined by the accuracy-through-order
procedure and are expressed through the coefficients $a_j$ of the
small-variable expansion (2), can be called, for short, the {\it root
approximant}. In the following sections, we demonstrate that this root
approximant provides quite accurate approximations for different problems,
uniformly extrapolating the small-variable expansion (2), valid for
$x \ra 0$, to the whole region of $x \in [0,\infty]$.

\section{Illustration by simple examples}

Before going to more complicated problems, we show the efficiency of the 
method by simple cases.

\subsection{Hard-core scattering problem}

Let us start the illustration of the method from the problem considered by
Baker and Gammel \cite{Baker_3}. When calculating the scattering length of a
repulsive square-well potential, one meets the integral
$$
 S(x) = \int_0^x \left ( \frac{\sin t}{t^3} \; - \; \frac{\cos t}{t^2}
\right )^2 \; dt \;  ,
$$
whose limit, as $x \ra \infty$, equals $\pi/15$. Baker and Gammel state that
this integral cannot be correctly evaluated by the standard Pad\'{e} method.
To solve the problem, they suggest a modified method employing a power of
the Pad\'{e} approximant. We show below that such integrals can easily be
treated by means of the root approximants.

The small-variable expansion of this integral reads as
$$
S(x) \simeq \frac{x}{9} \; - \; \frac{x^3}{135} + \frac{x^5}{2625} \; - \;
\frac{4x^7}{297675} + \frac{2x^9}{5893965} \; - \; \frac{x^{11}}{166080925} +
\frac{x^{13}}{10672286625} \;  .
$$
Comparing this with form (2), we have $S_0(x) = x/9$. Since expansion (2)
is in powers of $x^2$, we construct the root approximants (9) using $x^2$ as
a variable. Thus, the root approximant of third order is
$$
 S_3^*(x) = \frac{x}{9} \left ( \left ( \left ( 1 + A_1 x^2 \right )^2 +
A_2 x^4 \right )^{3/2} + A_3 x^6 \right )^{-1/6}\;  ,
$$
where the parameters are
$$
 A_1 = 0.133333 \; , \qquad A_2 = 0.012952 \; , \qquad A_3 = 0.016907 \;  .
$$
To fourth order,
$$
 S_4^*(x) = \frac{x}{9} \left ( \left ( \left ( \left ( 1 + A_1 x^2
\right )^2 + A_2 x^4 \right )^{3/2} + A_3 x^6 \right )^{4/3} +
A_4 x^8 \right )^{-1/8}\; ,
$$
where
$$
  A_1 = 0.133333 \; , \qquad A_2 = 0.012952 \; , \qquad A_3 = 0.002757 \; ,
\qquad A_4 = 0.004636 \;  .
$$
To fifth order,
$$
S_5^*(x) = \frac{x}{9} \left ( \left ( \left ( \left ( \left ( 1 + A_1 x^2
\right )^2 + A_2 x^4 \right )^{3/2} + A_3 x^6 \right )^{4/3} +
A_4 x^8 \right )^{5/4} + A_5 x^{10} \right )^{-1/10} \; ,
$$
where
$$
  A_1 = 0.133333 \; , \qquad A_2 = 0.012952 \; , \qquad A_3 = 0.002757 \; ,
$$
$$
 A_4 = 0.000578 \; , \qquad A_5 = 0.001285 \; .
$$
And to sixth order,
$$
S_6^*(x) = \frac{x}{9} \left ( \left ( \left ( \left ( \left ( \left (
1 + A_1 x^2 \right )^2 + A_2 x^4 \right )^{3/2} + A_3 x^6 \right )^{4/3} +
A_4 x^8 \right )^{5/4} + \right. \right.
$$
$$
\left. \left. +
A_5 x^{10} \right )^{6/5} + A_6 x^{12}
\right )^{-1/12} \;   ,
$$
where
$$
 A_1 = 0.133333 \; , \qquad A_2 = 0.012952 \; , \qquad A_3 = 0.002757 \; ,
$$
$$
A_4 = 0.000578 \; , \qquad A_5 = 0.000137 \; , \qquad A_6 = 0.000356 \;  .
$$
All these approximants converge to $\pi/15$, as $x \ra \infty$. The higher
the approximant order, the faster the convergence.

\subsection{Debye function}

The $n$-th order Debye function is defined \cite{Abramowitz_13} through the
integral representation
$$
 D(n,x) \equiv \frac{n}{x^n} \int_0^x \frac{t^n}{e^t -1 } \; dt \;  .
$$
For $|x| < 2 \pi$ and $n \geq 1$, it possesses the expansion
$$
 D(n,x) \simeq 1 \; - \; \frac{n}{2(n+1)} \; x \; + \;
n \sum_{k=1}^\infty \frac{B_{2k}}{(2k+n)(2k)!} \; x^{2k} \;  ,
$$
in which $B_{2k}$ are Bernoulli numbers. At large $x$ and ${\rm Re}\; n > 0$,
one has
$$
 D(n,x) \simeq \frac{C_n}{x^n} \qquad ( x\ra \infty , \; {\rm Re}\; n>0 ) \;  ,
$$
where
$$
C_n \equiv n \Gm(n+1) \zeta(n+1) \; .
$$

Below, we consider the case of $n = 3$, corresponding to the Debye function
$$
 D(x) \equiv D(3,x) = \frac{3}{x^3} \int_0^x \frac{t^3}{e^t-1} \; dt \;  .
$$
The small-variable expansion for the latter takes the form
$$
D(x) \simeq 1 - \; \frac{3}{8} \; x + \sum_{k=1}^\infty a_{2k} x^{2k}
\qquad ( x \ra 0 ) \; ,
$$
in which
$$
 a_{2k} = \frac{B_{2k}}{(2k+3)(2k)!} \;  .
$$
While the large-variable behavior is given by the expression
$$
D(x) \simeq   \frac{C_3}{x^3} \qquad ( x \ra \infty) \; ,
$$
with
$$
 C_3 = \frac{\pi^4}{5} = 19.481818 \;  .
$$

Constructing the root approximant
$$
 D_5^*(x) = \frac{x}{9} \left ( \left ( \left ( \left ( \left ( 1 + A_1 x
\right )^2 + A_2 x^2 \right )^{3/2} + A_3 x^3 \right )^{4/3} +
A_4 x^4 \right )^{5/4} + A_5 x^5 \right )^{-3/5} \;   ,
$$
we compare it with the exact numerical values of the function $D(x)$ and
find that $D^*_5(x)$ approximates well this function in the whole region of
$x \in [0,\infty]$, with the maximal error of $15 \%$ at $x = 5$. The best
two-point Pad\'{e} approximant of the same order, $P_{1/4}(x)$, is less
accurate, yielding the maximal error of $33 \%$ at $x = 15$.

\subsection{Fermi-Dirac integral}

The general form of the $j$-th order Fermi-Dirac integral is
$$
F(j,x) = \frac{1}{\Gm(j+1)} \int_0^\infty \frac{t^j}{e^{t-x}+1} \; dt \;   .
$$
Its asymptotic expansions are known \cite{Dingle_14}.

For concreteness, let us consider the zero-order case that reduces to the
function
$$
F(x) \equiv F(0,x) = \ln \left ( 1 + e^x \right ) \; .
$$
At small $x$, this function tends to $\ln 2$, and at large $x$, we have
$$
 F(x) \simeq x \qquad ( x \ra \infty ) \;  .
$$

The root approximant
$$
 F_5^*(x) = \ln 2 \left ( \left ( \left ( \left ( \left ( 1 + A_1 x
\right )^2 + A_2 x^2 \right )^{3/2} + A_3 x^3 \right )^{4/3} +
A_4 x^4 \right )^{5/4} + A_5 x^5 \right )^{1/5} \;   ,
$$
where
$$
   A_1 = 0.721348 \; , \qquad A_2 = 0.360674 \; , \qquad A_3 = 0.390257 \; ,
$$
$$
 A_4 = 0.410334 \; , \qquad A_5 = 4.294519 \;   ,
$$
provides an accurate approximation for the function $F(x)$ in the whole
region of $x \in [0,\infty]$, the maximal error being $5 \%$. The two-point
Pad\'{e} approximant $P_{3/2}(x)$ is slightly less accurate, with the
maximal error of $6 \%$.

\subsection{Fekete-Szeg\"o problem}

The problem of maximizing the absolute value of a functional in subclasses
of normalized functions is called the Fekete-Szeg\"o problem
\cite{Fekete_15,Dziok_16}. The Fekete-Szeg\"o functional is bounded by
the function
$$
 f(x) = 1 + 2 \exp \left ( - \; \frac{2x}{1-x} \right ) \;  ,
$$
where $0 < x < 1$.

In order to consider the interval $[0, \infty]$, as in other examples, we
can use the change of the variable
$$
 x = \frac{z}{1+z} \; , \qquad z = \frac{x}{1-x} \;  .
$$
Then $z \ra \infty$ as $x \ra 1$. Expanding $F(z)$ at small $z$ gives
$$
 F(z) \equiv f(x(z)) \simeq 3 - 4z + 4z^2 -\; \frac{8}{3} \; z^3 +
\frac{4}{3} \; z^4 \; - \; \frac{8}{15} \; z^5 \;  .
$$
The root approximant $F^*_3(z)$ uniformly approximates the function $F(z)$
on the interval $z \in [0,\infty]$, with the maximal error about $10 \%$.
The two-point Pad\'{e} approximant $P_{2/2}(z)$ is worse, having the maximal
error twice larger than the root approximant $F^*_3(z)$.

\section{Some useful tricks}

It is important to mention some tricks allowing for the convenient use of 
the method. Below, we discuss the interchange of small-variable and 
large -variable limits and the problem of dealing with logarithms.

\subsection{Inversion of expansions}

In the above examples, we have considered functions, whose expansions are 
better known for the small-variable limit, while a few, or just a single 
term, are available in the large-variable limit. But generally, the 
small-variable and large-variable limits are interchangeable. In those 
cases, when the large-variable expansion in powers of $1/x$ provides a 
number of terms and this expansion enjoys better convergence properties, 
it is possible to inverse the small-variable limit to the large-variable 
limit by using the variable change $x = 1/t$. Then, instead of the function 
$f(x)$, we consider the function
\be
\label{11}
 F(t) \equiv f \left ( \frac{1}{t} \right ) \; , \qquad
t = \frac{1}{x} \;  .
\ee

The small-variable limit (1) becomes the large-variable limit
\be
\label{12}
F(t) \simeq F^{(k)}(t) \equiv f_k\left ( \frac{1}{t} \right ) \qquad
( t \ra \infty ) \;  ,
\ee
in which
\be
\label{13}
F^{(k)}(t) = F_\infty(t) \left ( 1 +
\sum_{n=1}^k \frac{a_n}{t^n} \right ) \;   ,
\ee
with
$$
 F_\infty(t) \equiv f_0\left ( \frac{1}{t} \right ) = A t^{-\al} \;  .
$$

Conversely, the large-variable behavior (4) transforms to the small-variable
behavior
\be
\label{14}
 F(t) \simeq F_p(t) \equiv   f^{(p)} \left ( \frac{1}{t} \right ) \qquad
( t \ra 0 ) \; ,
\ee
in which
\be
\label{15}
 F_p(t) = F_0(t) \left ( 1 + \sum_{n=1}^p b_n t^n \right ) \;  ,
\ee
where
$$
F_0(t) \equiv f_\infty \left ( \frac{1}{t} \right )  = B t^{-\bt} \;  .
$$

After this change of the variable it is straightforward to employ the same
procedure of constructing the root approximants, as is explained in Sec. 2.

More generally, it is possible to use the change of the variable $t = 1/x^s$,
with a positive power $s > 0$, so that again $t \ra \infty$, when $x \ra 0$.

\subsection{Example of inversion}

As an illustration of the inversion procedure, we give below a typical example,
discussing it rather briefly, since the whole method of constructing the root
approximants is the same as before.

Let us consider the partition function of the so-called zero-dimensional
oscillator, or the generating functional of zero-dimensional $\varphi^4$ field
theory, which is defined through the integral
$$
I(x) = \frac{1}{\sqrt{\pi}} \int_{-\infty}^\infty \exp \left ( - \vp^2
- x \vp^4 \right ) \; d\vp \;   ,
$$
where $x$ plays the role of a coupling parameter. In the weak-coupling limit,
one has \cite{Yukalov_17} the asymptotic expansion
$$
 I(x) \simeq 1 + \sum_{n=1}^\infty a_n x^n \qquad ( x \ra 0 ) \;  ,
$$
in which the coefficients are
$$
 a_n = \frac{(-1)^n}{\sqrt{\pi}\; n!} \;
\Gm\left ( 2n + \frac{1}{2} \right ) \;  .
$$
For instance
$$
 a_1 = -\; \frac{3}{4} \; , \qquad a_2 = \frac{105}{32} \; , \qquad
a_3 = - \; \frac{3465}{128} \;  ,
$$
and so on.

The strong-coupling expansion reads as
$$
I(x) \simeq 1.022765 \; x^{-1/4} - 0.345684\; x^{-3/4} +
0.127846 \; x^{-5/4} \qquad ( x \ra \infty) \;  .
$$
Here the strong-coupling expansion provides a number of terms. Moreover,
the absolute values of the coefficients in this expansion diminish with
increasing order, contrary to the coefficients $a_n$ in the weak-coupling
expansion, which grow as $n^n$ with increasing order $n$. This makes the
strong-coupling expansion more suitable for constructing root approximants.

Resorting to the change of the variable $x = 1/t^4$, we consider the function
$J(t) \equiv I(1/t^4)$ and follow the scheme of the previous section. We
define the root approximants $J^*_k(t)$ that give us the approximants
$I^*_k(x) = J^*_k(1/x^{1/4})$ for the sought function. Found in that way
approximant $I^*_3(x)$ has the maximal error of $5 \%$ for the whole range
of $x \in [0, \infty]$. For comparison, the Pad\'{e} approximant $P_{1/2}(x)$
has the maximal error of about $20 \%$, which is much less accurate.

\subsection{Dealing with logarithms}

It is worth paying attention to the problem of series involving logarithms, 
which often appear in physics applications. Such series do not yield any 
complication for the method of root approximants described here.There are 
two equivalent ways of treating such series. Thus, if a series contains the 
terms with $x^n$, $x^{n+1}$, and with $x^n \ln x$, then it is
admissible to consider as the terms of one order either those containing
$x^n$ and $x^n \ln x$ or the terms $x^{n+1}$ and $x^n \ln x$.

As an illustration, let us consider, e.g., the typical form of such a series
involving logarithms as that one arising in the Nambu-Iona Lasinio model
\cite{Kunihiro_18} and leading to the function
$$
 f(x) = x \left [ \sqrt{1 + x^2} \; - \; x^2 \ln\left (
\frac{1+\sqrt{1+x^2}}{x} \right ) \right ] \;  ,
$$
where $x$ plays the role of mass. At asymptotically small $x$, it follows
$$
f(x) \simeq x + \left ( \frac{1}{2} - \ln 2 + \ln x \right ) x^3 \qquad
(x \ra 0 ) \; .
$$
While at large $x$, one has
$$
 f(x) \simeq \frac{2}{3} - \frac{1}{5x^2} + \frac{3}{28x^4} \qquad
( x \ra \infty ) \;  .
$$
Keeping in mind the dependence of the last expansion on $1/x^2$, it is
convenient to use the variable $z = 1/x^2$. The root approximant, satisfying
the required limits, has the form
$$
 f_4^*(x) = \frac{2}{3} \left ( \left ( \left ( 1 + A_1 z \right )^2 +
A_2 z^2 \right )^{3/2}  + A_3 z^2 \ln ( 1 + z ) + A_4 z^3 \right )^{-1/6} \; ,
$$
with all parameters uniquely defined by the given expansions. This expression
approximates well the initial function $f(x)$, with the maximal error of $2 \%$
at $x \approx 2$. Contrary to this, the best Pad\'{e} approximant of the same
order has the error of $11 \%$ at $x \approx 1.5$.

\section{Ground-state energy of electron gas}

Important and not trivial problems arise when studying the properties of 
charged systems \cite{Loos_19,Cioslowski_20,Cioslowski_21}. Here we show how
our method works for the case of homogeneous electron systems.

\subsection{One-dimensional electron gas}

The Hartree-Fock part of the uniform electron energy is well known. The 
problem arises in calculating the {\it correlation energy}. The latter is 
usually presented in a reduced dimensionless form $\varepsilon(r_s)$ as 
a function of the Seitz radius $r_s$. High-density expansion for 
one-dimensional uniform electron gas \cite{Loos_22} corresponds to small 
$r_s$, when for the correlation energy one has
$$
 \ep(r_s) \simeq C + 0.00845 r_s \qquad (r_s \ra 0 ) \; ,  
$$
where
$$
C = -\; \frac{\pi^2}{360} = - 0.027416 \;   .
$$
The low-density expansion \cite{Loos_22} implies large $r_s$, when
$$
 \ep(r_s) \simeq \frac{b_1}{r_s} + \frac{b_2}{r_s^{3/2}} \qquad
(r_s \ra \infty) \;  ,
$$
where
$$
 b_1 = - \left ( \ln \sqrt{2\pi} \;  - \; \frac{3}{4} \right ) =
-0.168939 \; , \qquad b_2 = 0.359933 \; .
$$

The root approximant, enjoying the same expansions, but valid for arbitrary 
$r_s$ reads as
$$
 \ep_3^*(r_s) = -\; \frac{\pi^2}{360}\; \left ( \left ( ( 1+ A_1 r_s)^{3/2}
+ A_2 r_s^2 \right )^{5/4} + A_3 r_s^3 \right )^{-1/3} \;  ,
$$
with the parameters
$$
 A_1 = 0.493150 \; , \qquad A_2 = 0.056122 \; , \qquad
A_3 = 0.004274 \;  .
$$
Comparing the prediction of the root approximant with the data from diffusion 
Monte Carlo calculations \cite{Loos_22} in the interval $0 < r_s < 20$, we
find that the maximal error of $\varepsilon^*_3$ is $8\%$. Pad\'{e} 
approximants give the errors between $2\%$ and $10\%$. Thus, 
$P_{1/2}(\sqrt{r_s})$ has the error of $2\%$, while $P_{0/3}(\sqrt{r_s})$ has 
the maximal error of $10\%$. The Cioslowski interpolation method 
\cite{Cioslowski_23} results \cite{Loos_22} in a better accuracy of $1\%$. 
However, this method includes an additional parameter that is fitted from  
numerical Monte Carlo calculations. While our aim has been in constructing 
good approximations without fitting parameters, being based only on asymptotic
expansions. The principal importance of avoiding fitting parameters is crucial 
for those problems where no exact numerical data are available.

\subsection{Two-dimensional electron gas}

Correlation energy of a homogenous two-dimensional electron gas was studied 
in several articles, e.g., in Refs. 
\cite{Sim_24,Tanatar_25,Kwon_26,Attaccalite_27,Gori_28,Constantin_29,
Drummond_30,Loos_31}. In high-density limit (small $r_s$), the ground-state 
energy reads \cite{Loos_31} as
$$
E_0(r_s) \simeq \frac{c_{-2}}{r_s^2} + \frac{c_{-1}}{r_s} +
\ep(r_s) \qquad (r_s \ra 0) \; ,
$$
where the first two terms constitute the Hartree-Fock energy, with
$$
 c_{-2} = \frac{1}{2} \; , \qquad c_{-1} = -\; \frac{4\sqrt{2}}{3\pi} \;  .
$$
And the last term is the correlation energy
$$
 \ep(r_s) \simeq c_0 + c_1' r_s \ln r_s \qquad (r_s \ra 0) \;  ,
$$
with the coefficients
$$
 c_0 = -0.192495 \; , \qquad c_1' = -\sqrt{2} \left ( \frac{10}{3\pi}
- 1 \right ) = - 0.0863136 \;  .
$$

In the low-density limit (large $r_s$) the asymptotic expansion for 
the correlation energy can be written \cite{Kwon_26} as
$$
 \ep(r_s) \simeq \frac{b_1}{r_s} + \frac{b_2}{r_s^{3/2}} +
\frac{b_3}{r_s^2} \qquad (r_s \ra \infty) \;   ,
$$
where
$$
  b_1 = -0.472189 \; , \qquad b_2 = 0.4964 \; , \qquad 
  b_3 = 0.5297 \; .
$$

For intermediate $r_s$, there have been suggested 
\cite{Attaccalite_27,Gori_28,Drummond_30} several phenomenological
expressions with parameters fitted from Monte Carlo calculations. Thus,
Gori-Giorgi et al. \cite{Gori_28} suggested the form
$$
 \ep(r_s) = A_0 + \left ( B_0 r_s + C_0 r_s^2 + D_0 r_s^3
\right ) \; \ln \left ( 1 + 
\frac{1}{E_0 r_s + F_0 r_s^{3/2}+G_0 r_s^2 +H_0 r_s^3} 
\right ) \;  ,
$$
with the parameters
$$
 A_0 = - 0.1925 \; , \qquad B_1 = 0.0863136 \; , \qquad
C_0 = 0.057234 \; , \qquad D_0 = 0.003362896 \;  .
$$
$$
 E_0 = 1.0022 \; , \qquad F_0 = -0.02069 \; , \qquad
G_0 = 0.34 \; , \qquad H_0 = 0.01747 \;   .
$$
This expression can be used as a numerical result for estimating the 
accuracy of approximate analytic formulas.

The root approximant, satisfying all asymptotic expansions reads as
$$
 \ep_5^*(r_s) = \frac{b_1}{r_s}  \left ( \left ( \left ( 1 +
\frac{A_1}{\sqrt{r_s}} \right )^2 + \frac{A_2}{r_s} \right )^{3/2}
+ \frac{A_3}{r_s}\; \ln \left ( 1 + \frac{1}{\sqrt{r_s}} \right )
+  \frac{A_4}{r_s^{3/2}} + \frac{A_5}{r_s^2} \right )^{-1/2} \; ,
$$
where the parameters are
$$
 b_1 = - 0.472189 \; , \qquad A_1 = 0.700849 \; , \qquad
A_2 = 2.723702 \; ,
$$
$$
 A_3 = 10.792193 \; , \qquad A_4 = -5.764339 \; , \qquad
A_5 = 6.017150 \;  .
$$
The error of this approximant is about $5\%$.

\section{Systems with spherical symmetry}

Finite quantum systems often enjoy spherical symmetry. Below, we consider
two examples of such systems that are important for applications.

\subsection{Energy of harmonium atoms}

An $N$-electron harmonium atom is described by the Hamiltonian
$$
 \hat H = 
\frac{1}{2} \sum_{i=1}^N \left ( - \nabla_i^2 + \om^2 r_i^2 \right ) + 
\frac{1}{2} \sum_{i\neq j}^N \frac{1}{r_{ij}} \;  ,
$$
where dimensionless units are employed and 
$$
 r_i \equiv | \br_{i} | \; , \qquad 
r_{ij} \equiv | \br_i - \br_j | \;  .
$$
This Hamiltonian provides a rather realistic modeling of trapped ions,
quantum dots, and some other finite systems, such as atomic nuclei and
metallic grains \cite{Birman_32}. This is why the energy of harmonium
atoms has been intensively studied 
\cite{Cioslowski_33,Cioslowski_34,Cioslowski_35,Cioslowski_36,Cioslowski_37}.
Here we show that root approximants give a good approximation for the energy 
of such systems. We consider the ground-state energy of a two-electron 
harmonium.   

At a shallow harmonic potential, the energy can be expanded 
\cite{Cioslowski_23} in powers of $\omega$, so that
$$
 E(\om) \simeq E_k(\om) \qquad (\om\ra 0) \;  ,
$$
with the truncated series
$$
 E_k(\om) = \sum_{n=0}^k c_n \om^{(2+n)/3} \;  .
$$
For instance, to third order, we get
$$
E_3(\om) = c_0 \om^{2/3} + c_1 \om + c_2 \om^{4/3} \;  ,
$$
with the coefficients
$$
 c_0 = \frac{3}{2^{4/3}} = 1.19055 \; , \qquad 
c_1 = \frac{1}{2} \; \left (  3 + \sqrt{3} \right ) = 2.36603 \; , 
\qquad    c_2 = \frac{7}{36}\; 2^{-2/3} = 0.122492 \; .
$$

And for a rigid potential, the energy is approximated \cite{Cioslowski_23} 
as
$$
 E(\om) \simeq E^{(p)}(\om) \qquad (\om\ra\infty) \;  ,
$$
where
$$
  E^{(p)}(\om) = \sum_{n=0}^p b_n \om^{(2-n)/2} \;  .
$$
To fourth order, one has
$$
 E^{(4)}(\om) = b_0 \om + b_1 \om^{1/2} + b_2 + b_3 \om^{-1/2} \;  ,
$$
where 
$$
b_0 = 3  \; , \qquad 
b_1 = \sqrt{\frac{2}{\pi} } = 0.797885 \; , \qquad
b_2 = -\;\frac{2}{\pi}\; \left ( 1 - \; \frac{\pi}{2} + 
\ln 2 \right )  = - 0.077891 \; ,
$$
$$
 b_3 = \left ( \frac{2}{\pi}\right )^{3/2}\; \left [ 2 - 2G - \; 
\frac{3}{2}\; \pi + ( \pi + 3) \ln 2 + \frac{3}{2}\; ( \ln 2)^2 - \; 
\frac{\pi^2}{24} \right )  = 0.0112528 \;  ,
$$
with the Catalan constant
$$
 G \equiv \sum_{n=0}^\infty \frac{(-1)^n}{(2n+1)^2} =
0.91596559 \; .
$$

The root approximant, respecting all given small-$\omega$, as well as 
large-$\omega$ expansions, is
$$
 E_6^*(\om) = c_0 \om^{2/3} \left (  \left ( \left ( \left ( \left ( 
\left ( 1 + A_1 \om^{1/3} \right )^{1/2} + A_2 \om^{2/3} \right )^{3/4}
+ A_3 \om \right )^{5/6} + A_4 \om^{4/3} \right )^{7/8} + \right. \right.
$$
$$
\left. \left. +
A_5 \om^{5/3} \right )^{9/10} + A_6 \om^2 \right )^{1/6} \; ,
$$
with the parameters
$$
c_0 = 1.19055  \; , \qquad A_1 = 48.4532 \; , \qquad
A_2 = 564.108 \; ,
$$
$$
 A_3 = 1088.39  \; , \qquad A_4 = 1221.08 \; , \qquad
A_5 = 796.791 \; , \qquad A_6 = 256 \; .
$$
We estimate the accuracy of the root approximant comparing it with the 
numerical data from Ref. \cite{Matito_38} and find that its maximal error 
is only $0.9 \%$. Note that Pad\'{e} approximants cannot be used in the 
case of harmonium, since the small-variable and large-variable asymptotic 
expansions are incompatible.

\subsection{Energy of two-electron spherium}

The two-electron spherium is a system consisting of two electrons that 
are confined to the surface of a sphere of radius $R$. The ground-state 
energy of the system \cite{Cioslowski_23,Loos_39} possesses the small-radius
expansion
$$
 E(R) \simeq \frac{1}{R} + c_0 + c_1 R + c_2 R^2 + c_3 R^3 \qquad 
(R \ra 0) \;  ,
$$
in which
$$
 c_0 = 4\ln 2 - 3 = - 0.22741128 \; , \qquad
c_1 = 8(\ln 2)^2 - 40 \ln 2 + 24 = 0.11773689 \; ,
$$
$$ 
c_2 = -0.05027560 \; , \qquad c_3 = 0.01395783 \;  .
$$
The coefficients $c_2$ and $c_3$ can also be expressed in closed forms that,
however, are too much cumbersome \cite{Loos_39}, because of which we give here
only their numerical values. 

In the large-radius limit, the energy has the expansion
$$
E(R) \simeq \frac{1}{2R} + \frac{1}{2R^{3/2}} \; - \; \frac{1}{8R^2} \;
- \; \frac{1}{128 R^{5/2}} \qquad (R \ra \infty ) \;   .
$$

The root approximant can be writen in the form
$$
 E_5^*(R) = \frac{1}{R} + c_0  \left ( \left ( \left (
\left ( ( 1 + A_1 R)^{3/2} + A_2 R^2 \right )^{5/4} + A_3 R^3 \right )^{7/6}
+ A_4 R^4 \right )^{9/8} + A_5 R^5 \right )^{-1/5} \;  ,
$$
where
$$
 A_1 = 1.05188915 \; , \qquad A_2 = 0.56453530 \; , \qquad 
A_3 = 0.36000617 \; , 
$$
$$
A_4 = 0.12606787 \; , \qquad A_5 = 0.01946301 \; .
$$
Comparing this expression with numerical data \cite{Loos_39}, we find that
the maximal error occurs at $R = 20$, being only $0.1 \%$. The best Pad\'{e}
approximant $P_{5/5}(\sqrt{R})$ is much less accurate, having the maximal 
error, also at $R = 20$, but an order larger, $1.5 \%$.

\section{Discussion}

We have described a simple and general method for interpolating functions
between their small-variable and large-variable asymptotic expansions. The
method is based on the construction of self-similar root approximants
enjoying the general form
$$
f^*_k(x) = f_0(x) \left ( \left ( \left ( \ldots ( 1 + A_1 x )^{n_1} +
A_2 x^2 \right )^{n_2} + A_3 x^3 \right )^{n_3} + \ldots +
A_k x^k \right )^{n_k} \; .
$$
All parameters $A_i$ can be uniquely defined through the corresponding
asymptotic expansions. By changing the variable, it is easy to inverse the
expansions between the small-variable and large-variable limits.

Our aim has been to suggest a method that would involve no fitting 
parameters. This is especially important in those complicated cases, where
numerical data in the whole region of the variable are not available. The 
absence of fitting parameters makes our aproach different from other 
intrepolation methods, such as the Cioslowski method \cite{Cioslowski_23}.  

We have demonstrated the method of root approximants by several examples,
whose structure is typical for many applications, including the hard-core
scattering problem, Debye function, Fermi-Dirac integral, Fekete-Szeg\"{o} 
problem, zero-dimensional oscillator, homogeneous electron gas, harmonium 
atom, and spherium.

We have analyzed several more problems, e.g., the interpolation of the 
polaron mass between weak-coupling and strong-coupling limits studied earlier 
by the Feynman variational procedure \cite{Feynman_39} and by other methods
\cite{Feranchuk_40,Alexandrou_41,Kleinert_42,Kornilovitch_43}. Our approach
provides approximations, whose accuracy is comparable or better than that
of other methods, being at the same time more simple.

Generally, the suggested method provides the accuracy not worse than the 
method of Pad\'{e} approximants and in the majority of cases is more accurate 
than the latter.

Except the root approximants of the general form (9), we also have considered 
{\it additive approximants} represented by the sums
$$
 f^*_{M/N}(x) = \sum_{i=1}^{(M+N)/2} A_i ( 1 + B_i x)^{n_i} \;  .
$$
This type of expressions can be considered either as additive root approximants 
or an additive variant resulting from self-similar factor approximants
\cite{Gluzman_44}.

For example, in the case of one-dimensional electron gas, the correlation
energy is approximated as
$$
\ep^*_{2/2}(r_s) = A_1 ( 1 + B_1 r_s )^{-1} +
 A_2 ( 1 + B_2 r_s )^{-3/2} \;  ,
$$
with the parameters
$$
 A_1 = -0.044941 \; , \qquad A_2 = 0.017526 \; , \qquad
B_1 = 0.266023 \; , \qquad B_2 = 0.133344 \;  .
$$
This expression has the maximal error of $11\%$. However a more detailed 
analysis of such additive approximants requires a separate investigation,
which is out of the scope of the present paper. 

\vskip 2cm

\acknowledgments{Acknowledgments}

One of the authors (V.I.Y.) acknowledges financial support from the Russian
Foundation for Basic Research (grant 14-02-00723) and is grateful for useful
discussions to E.P. Yukalova.

\conflictofinterests{Conflicts of Interest}

The authors declare no conflict of interest.

\end{document}